# *The Probability Sample in the Big Data Era: Theoretical Considerations and Practical Insights*

**Alessandro La Rocca**[1]

[1] *Italian National Institute of Statistics (Istat)*
*alessandro.larocca@istat.it*

## Abstract

This paper draws inspiration from the growing attention that national statistical offices are devoting to non-probabilitysamples and the possibility of integrating them with other sources, such as administrative registers and probability samples. Following a brief introduction on current trends related to new technologies and the wide availability of data, the contribution illustrates the main methodological approaches in the literature for the joint use of non-probability and probability samples. This integration aims to overcome the inferential limitations arising from the absence of a probability sampling frame.

## 1. Introduction

Non-probabilitysamples have attracted increasing interest in recent years. This interest is not confined to applied disciplines, such as economics and sociology, which have traditionally made extensive use of them because the phenomena under study do not allow data randomization for practical, and sometimes ethical, reasons; it also involves and above all the community of statistical methodologists. Indeed, the latter have intensified their scientific contributions aimed at developing methodologically sound estimation strategies capable of addressing the inferential challenges associated with the absence of a probability sampling design. The growing methodological debate, particularly from an inferential perspective, can be attributed to two main interrelated factors: technological progress in general, and in particular the data deluge (The Economist, “The data deluge”, 25 February 2010) made possible by the advent of the Internet.

The vast amount of data now available big data has characteristics that make it widely used by data scientists, notably its comprehensiveness, timeliness and heterogeneity. These are Doug Laney’s three “Vs”: Volume, Velocity and Variety (Laney, 2001). Web surveys and the availability of large administrative databases held by public authorities have also contributed to the use of non-probabilitysamples.

Official statistics are undergoing a major transition, moving from the solid but isolated framework of probability surveys towards an integrated ecosystem of heterogeneous data sources. Although probability sampling has provided unbiased design-based inference for decades, the current operational context is increasingly challenging its sustainability. Integrating non-probability data sources is no longer merely an option, but a strategic necessity driven by five critical factors:

Declining response rates: Lower participation not only reduces sample size but also introduces a potentially substantial systematic risk of nonresponse bias, shifting the burden of validity from the sampling design to statistical modelling.

Survey costs: The financial burden of traditional data collection networks has become prohibitive in relation to the rate at which data become obsolete.

Respondent burden: The length and complexity of institutional questionnaires generate survey fatigue, undermining the intrinsic quality of responses.

Demand for timeliness (real time statistics): Policy makers require granular and immediate evidence, at a pace that traditional production cycles cannot meet.

The big data explosion: The availability of web surveys and administrative data offers potentially very broad coverage; however, new inferential paradigms are needed to address selection bias.

This scenario calls for a shift from design based inference to model based inference, where the challenge lies in formalizing frameworks capable of handling non-probabilitydata with mathematical rigour.

## 2. The concept of representativeness

Conceptually, representativeness characterizes probability samples, just as selectivity characterizes non-probabilitysamples: in the former, inclusion is governed by a random design, whereas in the latter it results from a potentially informative selection mechanism. These are two key concepts that need to be clarified before examining estimation issues relating to non-probabilitysamples. A sample is said to be representative if the distribution of the characteristics of interest in the sample is the same as that in the reference population (Cochran, 1977). Here, representativeness is understood as equality between the sample and population distributions with respect to relevant variables. Groves et al. (2009) offer a less categorical and more gradual definition, describing representativeness as the extent to which a sample reflects the target population. Accordingly, a sample can be more or less representative. Meng (2018), instead, focuses not on a property of the sample, but on the data generation and selection mechanism. This definition is particularly relevant today because it emphasizes the data generation and selection mechanism, which is crucial for non-probabilitysamples. A probability sample is inherently representative of the population from which it is drawn in the sense that each unit has a known and positive probability of being included in the sample (inclusion probability), determined by the sampling design. This property provides a sound theoretical basis for statistical inference and, under appropriate conditions, makes it possible to obtain unbiased estimators of population parameters.

Conversely, in non-probabilitysamples, unit selection is generally driven by self selection or voluntary participation, so inclusion probabilities are unknown or cannot be defined (Bethlehem, 2010). Consequently, without further assumptions about the selection mechanism, unbiased estimation of the parameters of interest cannot be guaranteed, since selection is not random and often depends, directly or indirectly, on observed or unobserved unit characteristics. Non-probabilitysamples are subject to systematic distortions that can be summarized as follows: selection bias, whereby the inclusion mechanism favours certain groups; participation bias, associated with voluntary participation; and undercoverage bias, whereby entire segments of the population are excluded a priori.

The following sections describe the main methods proposed in the literature for obtaining unbiased or asymptotically unbiased estimates from non-probabilitysamples, under specific assumptions about the selection mechanism and the relationship between auxiliary variables and the variable of interest. Particular attention is paid to classical statistical methods, while more recent extensions based on machine learning techniques are not covered. The perspective adopted in this paper, widely shared in the literature, is to regard a non-probabilitysample as a special case of a probability sample affected by total nonresponse (Särndal et al., 1992; Valliant et al., 2013). In this setting, the subset of nonresponding units can be interpreted as the outcome of an informative selection mechanism which, under appropriate ignorability assumptions in the sense of Rubin (1976), requires sampling weight adjustments based on the inverse of inclusion probabilities in order to obtain unbiased estimates of population parameters. From this perspective, this paper does not address the ex ante assessment of whether selective forces are present in the non-probabilitysample. Although the assumption of ignorable selection is central to constructing unbiased estimators, in practice it is rarely verifiable and difficult to test using observed data alone (Bethlehem, 2010).

## 2.1 Estimation from non-probability samples

Many estimation methods currently used with non-probabilitysamples originate in techniques developed for probability surveys, particularly sampling weight adjustments for total nonresponse (Särndal et al., 1992; Little and Rubin, 2019). These approaches were subsequently adapted and generalized to non-probabilitysamples, where the absence of a known sampling design raises similar, albeit more pronounced, issues related to informative selection (Valliant et al., 2013). Even when a probability reference sample or administrative records are available, inference based on a non-probabilitysample requires assuming that the selection mechanism is ignorable conditional on the observed auxiliary variables. Reference data make it possible to estimate the population distribution of covariates and construct appropriate weight adjustments, but they do not eliminate the need for the ignorability assumption, which remains essential for identifying the parameter of interest (Chen et al., 2020). If the selection mechanism depends directly on the variable of interest, even conditional on observed covariates, selection is non-ignorable (NMAR). In this setting, as in non-ignorable total nonresponse in probability surveys, selection bias cannot be eliminated using weighting or modelling techniques based on observed data and must be addressed through assumptions or sensitivity analysis (Rubin, 1976; Little and Rubin, 2019; Bethlehem, 2010).

Sensitivity analysis examines the robustness of inference to controlled changes in unidentifiable assumptions about the selection or response mechanism. For example, a selection parameter may be varied over a specified range to assess how the conditional probability of sample inclusion, expressed through a logit model involving S (the sample), X (the covariate) and Y (the variable of interest), changes under ignorable (MAR) and non-ignorable (NMAR) assumptions.

## 3. Estimation methods

Estimation methods for non-probability samples can be grouped into three broad classes: model based estimators, pseudo weight estimators, and hybrid estimators. The first rely on correct specification of the outcome model, the second on correct modelling of the selection mechanism, while the third seek to combine both sources of information to improve robustness.

Let $P$ denote the population and $N$ its size, and let $S_p$ and $S_{np}$ denote the probability and non-probability samples, with sizes $n_p$ and $n_{np}$, respectively. Let $S$ be the union of two samples $S_p \cup S_{np}$, of size $n_p + n_{np}$. Let $Y_i$ denote the variable of interest and $X_i$ the covariates for $i = 1, \dots, n_p$ for $S_p$; for $i = 1, \dots, n_{np}$ for $S_{np}$. Let $\hat{\pi}_i$ denote the response propensity, and let $w_i$ and $d_i$ represent, respectively, the probability sample design weights and the weights associated with the non-probability sample.

## 3.1 Model-based estimators

Model-based estimators assume that the relationship between the variable of interest and the covariates is correctly described by a parametric or semiparametric statistical model. In particular, they assume that the relationship between the variable of interest $Y$ and a set of observed covariates $X$ can be represented by an appropriately specified statistical model. Formally, it is assumed that there exists a function $\mathrm{m}(.)$ such that:

$$E(Y|X) = m(X; \theta).$$

Here, $\mathrm{m}(.)$ is a known function (e.g. linear or logistic), and $\theta$ is a vector of unknown parameters to be estimated.

In non-probability samples, these methods are typically implemented by estimating the outcome model using the non-probability sample, where $Y$ is observed, and then using the fitted model to predict values of the variable of interest for a probability reference sample or for the entire population. This procedure is known in the literature as mass imputation (Rivers, 2007; Kim and Rao, 2012). If a probability reference sample is available $S_p$, with design weights $w_i$ and pseudo weights $d_i = 1/\pi_i$ based on the estimated propensity $\pi_i$, the mean estimator can be expressed as follows:

$$\hat{\mu}_{reg} = \frac{1}{\sum_{i\epsilon S_p} w_i} \sum_{i\epsilon S_p} w_i \, \widehat{m}(x_i).$$

Here, $\widehat{m}(x_i)$ the predicted values are obtained from the fitted model. The efficiency of these estimators critically depends on correct specification of the outcome model. In the event of misspecification, bias may be substantial and persist even as sample size increases (Chambers and Clark, 2012).

## 3.2 Pseudo-weighting or quasi-randomization estimators

Quasi-randomization methods are often referred to in the literature as pseudo-weighting methods because they rely on estimated weights that mimic design weights without being derived from an actual probability sampling design. These methods seek to correct selectivity in non-probability samples by estimating, ex post, the units' inclusion probabilities. The resulting weights, defined as the reciprocals of estimated propensities, do not arise from a known sampling design and therefore cannot be interpreted as classical design weights; rather, they are pseudo-weights used for adjustment purposes. Quasi-randomization explicitly models the mechanism by which units are selected into the non-probability sample, rather than modelling the variable of interest. A propensity score is introduced and defined as follows:

$$\hat{\pi}_i(X) = P(S_i = 1|X).$$

Under the assumption of ignorable selection conditional on the covariates:

$$S \perp Y|X,$$

units in the non-probability sample can be weighted using pseudo-weights:

$$d_i = \frac{1}{\widehat{\pi_i}}$$

these methods make it possible to treat the non-probability sample as approximately equivalent to a probability sample, provided the overlap assumption holds, namely:

$$0 < \pi_i(X_i) < 1.$$

For all relevant covariate configurations (Rosenbaum and Rubin, 1983; Bethlehem, 2010). The estimator of the mean adopted here is the normalized Hájek estimator, which is more stable than the Horvitz-Thompson estimator and has lower variance.

$$\hat{\mu}_{PS} = \frac{\sum_{i \epsilon S} \frac{S_i y_i}{\widehat{\pi_i}}}{\sum_{i \epsilon S} \frac{S_i}{\widehat{\pi_i}}}.$$

Overlap (or common support) refers to the condition that, conditional on observed covariates, units in the population can be selected into either the probability or non-probability sample. In the absence of overlap, some regions of the covariate space are observed exclusively in one of the two samples, making identification of estimators based on weighting or selection models impossible. When overlap is present, units with similar covariate profiles may belong to both samples, allowing quasi-randomization procedures to be applied. Conversely, lack of overlap is a structural limitation that prevents estimator identification. Formally, this condition can be expressed as:

$$0 < P(S = 1|X) < 1 \text{ for all } X,$$

model-based and pseudo-weight estimators have complementary strengths and weaknesses. Model-based estimators can be efficient but depend heavily on correct specification of the outcome model; the generation of extreme weights is also a limitation. Pseudo-weight estimators are less dependent on the functional form of the outcome model, but may be unstable when estimated propensities are extreme or overlap is limited.

## 3.3 Doubly robust estimators

Doubly robust (DR) estimators are a class of procedures combining a model for the variable of interest with a model for the selection mechanism, with the aim of improving inferential robustness when dealing with non-probability samples or nonresponse. The underlying idea is to exploit information from both an outcome model and a selection model, providing protection against misspecification of either model (Robins, Rotnitzky and Zhao, 1994; Bang and Robins, 2005). A doubly robust estimator of the population mean can be expressed as:

$$\hat{\mu}_{DR} = \frac{\sum_{j \epsilon S_p} w_j \widehat{m}\left(x_j\right)}{\sum_{j \in S_p} w_j} + \frac{\sum_{i \in S_{np}} \hat{d}_i \left(Y_i - \widehat{m}(x_i)\right)}{\sum_{i \in S_{np}} d_i},$$

the first term is a model-based estimate (mass imputation), obtained by predicting the variable of interest in the probability sample, while the second term introduces a correction based on residuals observed in the non-probability sample, appropriately weighted by pseudo-weights. The estimator's structure shows that the weighting component acts only on the part not explained by the outcome model, thereby reducing the impact of potential model misspecification. The fundamental property of DR estimators is double robustness: the estimator is consistent if at least one of the outcome $m(X)$ or selection models $\pi(X)$ is correctly specified. Specifically, if the outcome model is correct, the estimator is consistent regardless of whether the propensity score model is correct; conversely, if the selection model is correctly specified, consistency is guaranteed even if the outcome model is misspecified. If both models are correct, the DR estimator generally achieves greater efficiency than estimators relying on a single model (Bang and Robins, 2005; Kang and Schafer, 2007). Despite these properties, DR estimators have limitations. In particular, they remain sensitive to limited overlap and estimated propensities close to zero, which can lead to numerical instability and high variance. Moreover, double robustness does not protect against cases in which both models are severely misspecified or the selection mechanism depends directly on the variable of interest in a non-ignorable manner (NMAR).

## 3.4 Kernel estimators

In inference from non-probability samples, kernel estimators fall within the quasi-randomization framework, since they construct pseudo-probability weights for units in the non-probability sample using a probability reference sample. This methodological strand developed as a nonparametric extension of propensity-score-based approaches (Rosenbaum and Rubin, 1983) and has recently been adapted to the integration of heterogeneous data sources (probability and non-probability), as discussed, for example, by Chen, Li and Wu (2020) and Yang and Kim (2020). Formally:

$$d_i^{kw} = \sum_{j \in S_p} w_j \, K_h\left(\left\| X_i - X_j \right\|\right).$$

Alternatively, if computed on propensity scores:

$$d_i^{kw} = \sum_{j \in S_p} w_j \, K_h\left(\hat{\pi}_i - \hat{\pi}_j\right),$$

the estimator of the mean is then:

$$\hat{\mu}_{kw} = \frac{\sum_{i \epsilon S_{np}} d_i^{kw} Y_i}{\sum_{i \epsilon S_{np}} d_i^{kw}},$$

here, $K_h(.)$ is a kernel function and h is the smoothing parameter (Wand and Jones, 1995). The basic idea is to use the probability sample as a representation of the reference covariate distribution, distributing design weights across units in the non-probability sample according to their proximity in covariate space or propensity-score space (Chen et al., 2020). Compared with model-based approaches (such as mass imputation), which rely on specifying a model for the outcome $E(Y|X)$ (Kim and Rao, 2012), and pseudo-weight approaches based on inverse inclusion probabilities Rosenbaum and Rubin, 1983), kernel estimators provide an intermediate solution. They are generally less sensitive to misspecification of the selection model than pseudo-weight estimators and, at the same time, avoid the exclusive reliance on an outcome model characteristic of model-based approaches. In this sense, they can be interpreted as nonparametric model-assisted estimators (Deville and Särndal, 1992).

Recent developments have also shown that kernel estimators can be combined with outcome models in a doubly robust framework, achieving consistency when at least one of the selection or outcome models is correctly specified (Bang and Robins, 2005; Chen et al., 2020). With adequate overlap between samples, these methods offer a flexible and robust approach to correcting selection bias in non-probability samples (Bethlehem, 2010).

## 3.5 Model-assisted kernel estimators

The Model-Assisted Kernel Weighting (MAKW) estimator lies at the intersection of model-based and quasi-randomization approaches, combining an outcome regression model with a nonparametric kernel-based weight calibration procedure. In analogy with model-assisted estimators in sampling theory (see, for example, Särndal, Swensson and Wretman, 1992), MAKW uses a regression model estimated from the probability sample to reconstruct the systematic component of the variable of interest in the population, while the residual component is adjusted through kernel weights constructed from the non-probability sample, with the aim of restoring representativeness with respect to observed covariates. This framework follows the logic of calibration and generalized regression estimation methods (Deville and Särndal, 1992), but replaces linear calibration with a distance-based (or propensity-score-based) nonparametric smoothing function, as discussed in the literature on kernel weighting estimators for non-probability samples (e.g. Chen, Li and Wu, 2020). MAKW can therefore be interpreted as a hybrid estimator combining robustness to outcome-model specification with the nonparametric flexibility of kernel weights, making it particularly useful when the relationship between covariates and the selection mechanism is complex or nonlinear (Cobo et al., 2025). Compared with estimators based solely on propensity scores or pure kernel estimators, MAKW may reduce bias under moderate model misspecification and provide better variance control, representing a natural model-assisted extension of approaches for integrating probability and non-probability samples.

$$\hat{\mu}_{makw} = \frac{1}{N} \sum_{i \in S_p} w_i \, \hat{m}(x_i) + \frac{1}{N} \sum_{i \in S_{np}} w_i^{kw} \left( y_i - \hat{m}(x_i) \right)$$

$$w_i^{kw} = \frac{\sum_{j \in S_p} w_j \, k_h\left( \left\| x_i - x_j \right\| \right)}{\sum_{j \in S_{np}} k_h\left( \left\| x_i - x_j \right\| \right)}$$

## 3.6 Kernel doubly robust estimators

The Kernel Doubly Robust (KDR) estimator can be interpreted as a natural nonparametric extension of doubly robust (DR) estimators for integrating probability and non-probability samples. In line with the literature on inverse probability weighting (IPW) and propensity score adjustment, the estimator exploits the fact that the ratio of the inclusion densities of the two samples can be expressed as a function of the propensity score (Rosenbaum and Rubin, 1983; Little and Rubin, 2019). Rather than modelling the propensity parametrically, however, KDR uses it as a one-dimensional smoothing variable, estimating the density ratio nonparametrically through a kernel, similarly to density-ratio estimation approaches (Hirano, Imbens and Ridder, 2003). The estimator can thus be written as follows:

$$\mu^{kdr} = \frac{1}{N}\sum_{i\in S_p} w_i\, \widehat{m}(x_i) + \frac{1}{N}\sum_{i\in S_{np}} r_i^{kdr}\,(\widehat{\pi}_i)\big(y_i - \widehat{m}(x_i)\big),$$

here:

$$r_i^{kdr} = \frac{\sum_{j\in S_p} w_j\, k_h\big(\widehat{\pi}_i - \widehat{\pi}_j\big)}{\sum_{j\in S_{np}} k_h\,\big(\widehat{\pi}_i - \widehat{\pi}_j\big)}.$$

The first component is a model-assisted term (Särndal, Swensson and Wretman, 1992), while the second is a weighted correction that approximates the selection mechanism using a kernel over the propensity-score space. Like classical DR estimators (Bang and Robins, 2005; Chen, Li and Wu, 2020), KDR is consistent if at least one of the outcome or selection models is correctly specified. Its advantage is that it avoids parametric misspecification of the propensity model and reduces the dimensionality of the problem, mitigating the curse of dimensionality typically encountered with kernel estimators over covariates (Wand and Jones, 1995). This construction makes it particularly suitable for non-probability samples treated as missing at random (MAR), where the inclusion mechanism depends on observed covariates but is not known in parametric form (Elliott and Valliant, 2017). The estimator uses the kernel in propensity-score space to approximate the density ratio between the two samples. Formally:

$$\frac{f_p(\pi)}{f_{np}(\pi)}$$

## 4. Concluding Remarks

The rapid expansion of non-probability data sources and big data is reshaping the paradigm of official statistics. While traditional probability sampling remains the gold standard for unbiased design-based inference, declining response rates, rising costs, and the demand for timely indicators make data integration an unavoidable strategic priority.

As reviewed in this paper, the statistical literature provides a well-established toolkit ranging from mass imputation and inverse propensity pseudo-weighting to doubly robust and nonparametric kernel estimators to mitigate selection bias when integrating non-probability samples with probability reference surveys. Hybrid approaches, such as Doubly Robust (DR), Model-Assisted Kernel Weighting (MAKW), and Kernel Doubly Robust (KDR) estimators, are particularly promising because they combine outcome modeling with flexible selection adjustments, offering protection against partial model misspecification.

However, no estimation strategy can fully compensate for structural deficiencies in the data-generating process. The validity of all discussed estimators critically hinges on two foundational assumptions: conditional ignorability (MAR) given the observed auxiliary variables and adequate covariate overlap (common support) between the probability and non-probability samples. When selection directly depends on unobserved characteristics or on the study variable itself (NMAR), standard weighting and imputation techniques fail to eliminate bias. Future developments in official statistics should therefore pair these estimators with systematic sensitivity analyses for non-ignorable selection, alongside machine-learning extensions capable of capturing complex, high-dimensional relationships between covariates, selection mechanisms, and target variables.

## References

Bang, H., & Robins, J. M. (2005). Doubly robust estimation in missing data and causal inference models. *Biometrics*.

Bethlehem, J. (2010). Selection bias in web surveys. *International Statistical Review*.

Cattaneo, M. D., Jansson, M., & Ma, X. (2020). Simple local polynomial density estimators. *Journal of the American Statistical Association*.

Chambers, R. L., & Clark, S. J. (2012). *An Introduction to Model-Based Survey Sampling with Applications*. Oxford University Press.

Chen, Y., Li, P., & Wu, C. (2020). Doubly robust inference with non-probability survey samples. *Journal of the American Statistical Association*.

Cobo, B., Rueda-Sánchez, J. L., Ferri-García, R., & Rueda, M. M. (2025). A new technique for handling non-probability samples based on model-assisted kernel weighting. *Mathematics and Computers in Simulation*.

Cochran, W. G. (1977). *Sampling Techniques*. Wiley.

Deville, J.-C., & Särndal, C.-E. (1992). Calibration estimators in survey sampling. *Journal of the American Statistical Association*.

Elliott, M. R., & Valliant, R. (2017). Inference for nonprobability samples. *Statistical Science*.

Groves, R. M., et al. (2009). *Survey Methodology*. Wiley.

Härdle, W. (1990). *Applied Nonparametric Regression*. Cambridge University Press.

Hirano, K., Imbens, G. W., & Ridder, G. (2003). Efficient estimation of average treatment effects using the estimated propensity score. *Econometrica*.

Kang, J. D., & Schafer, J. L. (2007). Demystifying double robustness: A comparison of alternative strategies for estimating a population mean from incomplete data. *Statistical Science*.

Kim, J. K., & Rao, J. N. K. (2012). Combining data from two independent surveys: A unified approach. *Biometrika*.

Laney, D. (2001). *3D Data Management: Controlling Data Volume, Velocity, and Variety*. META Group Research Note.

Little, R. J. A., & Rubin, D. B. (2019). *Statistical Analysis with Missing Data* (3rd ed.). Wiley.

Meng, X.-L. (2018). Statistical paradises and paradoxes in big data (I): Law of large populations, big data paradox, and self-correction index. *Annals of Applied Statistics*.

Rivers, D. (2007). Sampling from non-probability panels. *Proceedings of the Survey Research Methods Section*.

Robins, J. M., Rotnitzky, A., & Zhao, L. P. (1994). Estimation of regression coefficients when some regressors are not always observed. *Journal of the American Statistical Association*.

Rosenbaum, P. R., & Rubin, D. B. (1983). The central role of the propensity score in observational studies for causal effects. *Biometrika*.

Rubin, D. B. (1976). Inference and missing data. *Biometrika*.

Särndal, C.-E., Swensson, B., & Wretman, J. (1992). *Model Assisted Survey Sampling*. Springer-Verlag.

Sugiyama, M., et al. (2012). *Density Ratio Estimation in Machine Learning*. Cambridge University Press.

Valliant, R., et al. (2013). *Practical Tools for Designing and Weighting Survey Samples*. Springer.

Wand, M. P., & Jones, M. C. (1995). *Kernel Smoothing*. Chapman & Hall.

Yang, S., & Kim, J. K. (2020). Statistical data integration in survey sampling: A review. *Japanese Journal of Statistics and Data Science.*